\documentclass[%
twocolumn,
superscriptaddress,
nofootinbib,
 amsmath,amssymb,
 aps, prd
]{revtex4-2}

\usepackage{graphicx}
\usepackage{tensor}
\usepackage{siunitx}
\usepackage{xcolor} 
\usepackage[normalem]{ulem} 
\usepackage{lipsum} 

\usepackage[pdfborder={0 0 0}, plainpages=false, hyperfigures=true]{hyperref}

\newcommand{\beq}{\begin{equation}}
\newcommand{\eeq}{\end{equation}}
\newcommand{\beqn}{\begin{eqnarray}}
\newcommand{\eeqn}{\end{eqnarray}}

\newcommand{\lo}{\mathrel{\raise.3ex\hbox{$<$}\mkern-14mu
    \lower0.6ex\hbox{$\sim$}}}
\newcommand{\go}{\mathrel{\raise.3ex\hbox{$>$}\mkern-14mu
    \lower0.6ex\hbox{$\sim$}}}

\newcommand{\WSU}{\affiliation{Department of Physics \& Astronomy,
	Washington State University, Pullman, Washington 99164, USA}}
\newcommand{\UNH}{\affiliation{Department of Physics \& Astronomy, University of New Hampshire, 9 Library Way, Durham NH 03824, USA}}
\newcommand{\TAPIR}{\affiliation{TAPIR, Walter Burke Institute for Theoretical Physics, MC 350-17, California Institute of Technology, Pasadena, California 91125, USA}}
\newcommand{\Cornell}{\affiliation{Cornell Center for Astrophysics and Planetary Science, Cornell University, Ithaca, New York, 14853, USA}}
\newcommand{\MPI}{\affiliation{Max Planck Institute for Gravitational Physics (Albert Einstein Institute), Am M{\H u}hlenberg 1, 14476 Potsdam, Germany}}

\begin{document}
\title{Dynamical ejecta from binary neutron star mergers: Impact of residual eccentricity and equation of state implementation}
\author{Francois Foucart}\UNH
\author{Matthew D. Duez}\WSU
\author{Lawrence E. Kidder}\Cornell
\author{Harald P. Pfeiffer}\MPI
\author{Mark A. Scheel}\TAPIR

\begin{abstract}

Predicting the properties of the matter ejected during and after a neutron star merger is crucial to our ability to use electromagnetic observations of these mergers to constrain the masses of the neutron stars, the equation of state of dense matter, and the role of neutron star mergers in the enrichment of the Universe in heavy elements. Our ability to reliably provide such predictions is however limited by a broad range of factors, including the finite resolution of numerical simulations, their treatment of magnetic fields, neutrinos, and neutrino-matter interactions, and the approximate modeling of the equation of state of dense matter. In this manuscript, we study specifically the role that a small residual eccentricity and different implementations of the same equation of state have on the matter ejected during the merger of a $1.3M_\odot-1.4M_\odot$ binary neutron star system. We find that a residual eccentricity $e\sim 0.01$, as measured $\sim 4-6$ orbits before merger, causes $O(25\%-30\%)$ changes in the amount of ejected mass, mainly due to changes in the amount of matter ejected as a result of core bounces during merger. We note that $O(1\%)$ residual eccentricities have regularly been used in binary neutron star merger simulations as proxy for circular binaries, potentially creating an additional source of error in predictions for the mass of the dynamical ejecta.
  
\end{abstract}

\maketitle

\section{Introduction}

Over the last few years, neutron star mergers have become an important source of information for theoretical nuclear physics. The observation of the first binary neutron star merger, GW170817, through gravitational waves~\cite{TheLIGOScientific:2017qsa} provided us with constraints on the equation of state of neutron rich matter above nuclear saturation density~\cite{GW170817-NSRadius}. In that respect, gravitational wave observations complement measurements of neutron star radii by NICER~\cite{Miller:2021qha,Raaijmakers:2021uju}, astrophysical constraints on the maximum mass of neutron stars~\cite{Demorest:2010bx}, as well as laboratory measurements of heavy, neutron-rich nuclei~\cite{Horowitz:2013wha,PhysRevLett.126.172503}. The ejection of neutron rich matter during and after merger also provides us with important information about neutron stars, and their potential role in enriching the Universe in heavy elements. Neutron rich ejecta from binary neutron star and black hole-neutron star mergers undergoes r-process nucleosynthesis, and radioactive decay of the ashes of the r-process power a UV/optical/infrared signal hours to weeks after the merger -- a {\it kilonova}. Kilonovae carry information about the properties of the merging compact objects (including additional information about the equation of state of dense matter)~\cite{Coughlin:2018fis,Nedora:2020qtd}, and about the elements synthesized in the ejecta~\cite{2013ApJ...775...18B}.

Extracting information from kilonovae is a complex problem, requiring merger simulations accounting for general relativity, magnetic fields, and neutrino-matter interactions, as well as radiation-hydrodynamics simulation of the evolution of the ejected matter as it is continuously heated by nuclear reactions. As a result, our current ability to model kilonovae remains limited by both the accuracy of the simulations as well as the manner in which they incorporate (often approximately) important nuclear physics: dense matter equation of state, neutrino transport, neutrino-matter interactions, neutrino oscillations, out-of-equilibrium nuclear reactions, photon transport and opacities in the ejecta,... While these issues are well-known, their actual impact on our ability to extract information from kilonovae remains uncertain.

In this manuscript, we begin an investigation of the robustness of simulation results to changes in the simulation setup and the input nuclear physics. To avoid confusing the many potential sources of uncertainty affecting neutron star mergers, we begin here with general relativistic hydrodynamical simulations without magnetic fields nor neutrinos, and explore the sensitivity of the inspiral, merger and early matter ejection to small changes in the initial conditions (binary eccentricity), exact implementation of the dense matter equation of state, and numerical resolution. Previous simulations have already shown that large eccentricities can lead to the ejection of much larger amount of mass during tidal disruption of a neutron star (see e.g.~\cite{East:2015vix,Radice:2016}). Our results show that small changes O($1\%$) in the initial eccentricity of the binary can impact predictions for the ejected matter well above estimated numerical errors. This is an important consideration when attempting to predict the mass ejected by any given binary system, as many existing numerical relativity simulations have $O(1\%)$ residual eccentricity in their initial conditions. In addition, the simulations presented here can serve as a baseline for further investigation of the impact of nuclear inputs and/or numerical methods on the main observables extracted from merger simulations.

\section{Equations of State}
\label{sec:eos}

All simulations presented in this manuscript are performed with the SFHo equation of state~\cite{Steiner:2012rk}. We note however that different implementations of this equation of state are currently available, and that some post-processing is applied to publicly available equations of state in order to make them usable by numerical simulations. Specifically, our numerical simulations require subluminal sound speeds and a one-to-one map between physical variables (density, temperature, electron fraction) and evolved variables (conserved mass density, energy density, momentum density).

\begin{figure}
\includegraphics[width=0.9\columnwidth]{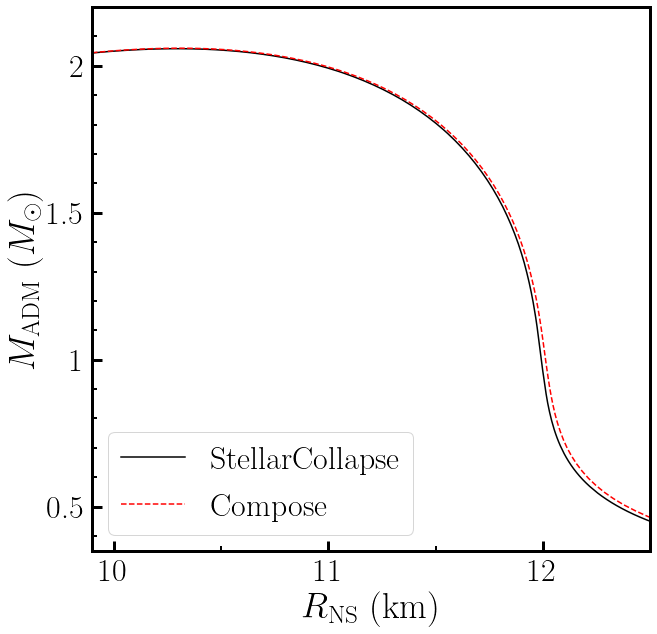}
\includegraphics[width=0.9\columnwidth]{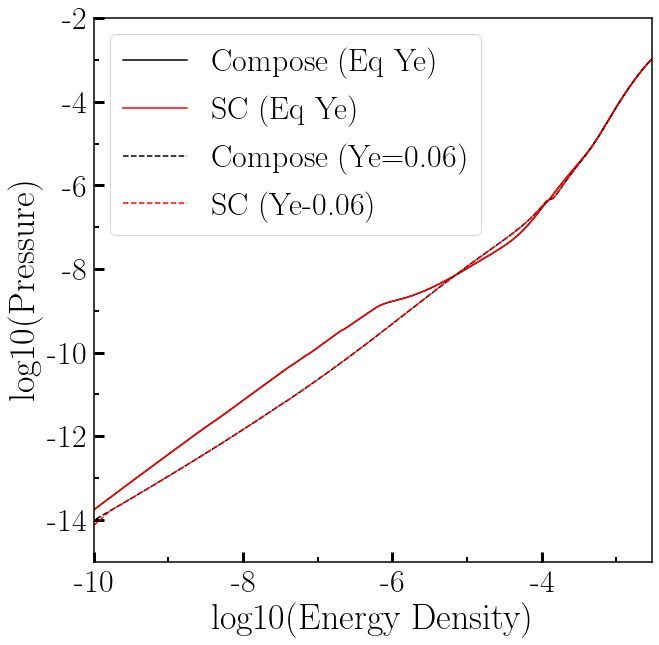}
\includegraphics[width=0.9\columnwidth]{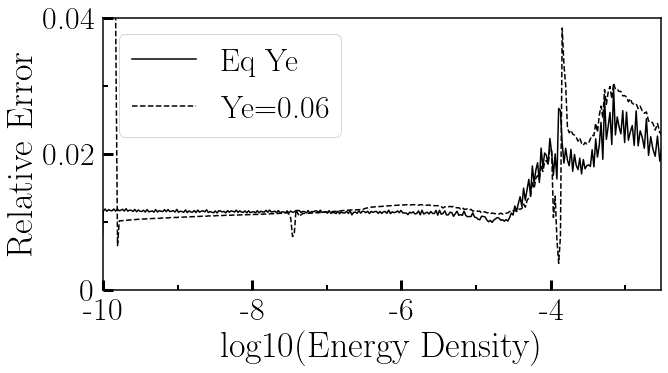}
 \caption{{\it Top}: Mass-Radius relationship for the two equations of state used in this manuscript. For the range of masses used in our simulations, the radii differ by $\sim 20\,{\rm m}$ or less. {\it Middle}: Pressure as a function of energy density for the same equations of state. We consider pressure at $T=0.1\,{\rm MeV}$ and for either the initial equilibrium $Y_e$ or for fixed $Y_e=0.06$. We show results in units of $G=M_\odot=c=1$, for the range of densities accessed by the simulations. {\it Bottom}: Relative error between the two equations of state, for the same quantities as in the middle panel.}
\label{fig:eos}
\end{figure}

\begin{table}
\begin{tabular}{c|c|c|c|c}
EoS Table & $M_{\rm ADM}$ & $M_b$ & $R_{\rm circ}$ & Simulations\\
\hline
Units & $M_\odot$ & $M_\odot$ & km & \\
\hline
SC & 1.30 & 1.44 & 11.92 & SC-MR/SC-HR \\
SC & 1.40 & 1.56 & 11.88 & SC-MR/SC-HR \\
Comp & 1.30 & 1.44 & 11.94 & Comp-no-ecc\\
Comp & 1.40 & 1.56 & 11.90 & Comp-no-ecc\\
Comp & 1.29 & 1.42 & 11.95 & Comp-ecc\\
Comp & 1.39 & 1.55 & 11.90 & Comp-ecc\\
\end{tabular}
\caption{Properties of neutron stars evolved in this manuscript.
The table from which we construct the SFHo equation of state (EoS) is either from StellarCollapse (SC)~\cite{OConnor2010} or Compose (Comp)~\cite{Typel:2013rza,Oertel:2016bki}, post-processed for use in SpEC simulations. We also quote the ADM mass, baryon mass (defined through $m=N_b m_n$ with $N_b$ the number of baryons and $m_n$ the neutron mass), and circular radius of each neutron star, as well as the simulations in which they were used (using the notation of Table~\ref{tab:sim}).}
\label{tab:ns}
\end{table}

To study the impact of these implementation details, we consider two distinct tabulated versions of the SFHo equation of state. First, we start from the table available in the Compose database~\cite{Typel:2013rza,Oertel:2016bki}, with a minimal amount of post-processing. We read the pressure $P$, specific internal energy $\epsilon$, and sound speed square $c_s^2$ from the table. The values are tabulated as functions of the logarithm of the baryon number density $n_b$, logaritum of the temperature $T$, and electron fraction $Y_e$. Whenever a point in the table has sound speed square $c_s^2>1$ or $c_s^2<0$, we correct the value of $(P,\epsilon,c_s^2)$ by simply taking the average value of all neighboring points in the table (in 3D; thus from 8 neighboring points). We additionally correct the pressure to guarantee that $dP/dT\geq 0$ everywhere, through linear interpolation of the logarithm of $P$ from neighboring point of the table (in 1D only here at constant density and electron fraction; thus from 2 neighboring points).

Our second version starts from the StellarCollapse~\cite{OConnor2010} table. As for the Compose equation of state table, we correct the pressure to guarantee that $dP/dT\geq 0$ everywhere. The two equations of state are close to each other (by the standard of the methods typically used to produce equations of state matching specific nuclear physics models), but not exactly identical. Table~\ref{tab:ns} shows the gravitational (ADM) mass, baryon mass, and radii of the stars evolved in this manuscript, while Fig.~\ref{fig:eos} shows the mass-radius relation of cold isolated neutron stars for both equations of state, the pressure for two one-dimensional slices through the table, and the relative differences in that pressure. We see differences of $O(1\%)$ or less between the two implementations. The two tables also differ in their definition of the baryon mass at the $O(1\%)$ level, due to different choices for the reference mass of a single baryon (see below). We note that while we will refer to these equations of state as `StellarCollapse' and `Compose' here, the differences between tables used in our simulations are likely due in part at least to interpolation from tables with different grids in density, temperature, and electron fraction -- i.e. they may tell us more about errors due to the finite resolution of the tables than about differences between StellarCollapse and Compose.

\section{Initial Data and Eccentricity}
\label{sec:id}

\begin{table}
\begin{tabular}{c|c|c|c}
Sim & EoS Table & $\Delta x_{0,\rm FD}$ [m] & $e_0$\\
\hline
SC-MR & SC & $ 205 $ & 0.003\\
SC-HR & SC & 165 & 0.003 \\
Comp-no-ecc & Comp & 205 & 0.004\\
Comp-ecc & Comp & 205 & 0.01 \\
\end{tabular}
\caption{List of simulations performed for this manuscript. The equation of state (EoS) table is defined as on Table~\ref{tab:ns}. The initial grid spacing of the finite difference grid $\Delta x_{0,\rm FD}$ and initial eccentricity $e_0$ are also provided. Note that in the coordinates of the simulation (quasi-isotropic coordinate), the radius of each neutron star is only $\sim 9.3\,{\rm km}$; additionally, the grid resolution at later times is always finer than the initial resolution (by as much as $\sim 20\%$).}
\label{tab:sim}
\end{table}

We generate initial conditions using the Spells initial data solver~\cite{Pfeiffer2003,FoucartEtAl:2008}. Spells begins by generating quasi-equilibrium binaries with zero initial radial velocity (``quasi-circular'' binaries) for user-provided values of the neutron star baryon masses and binary separation. This typically results in orbits with residual eccentricity $e\sim 0.01$, which can be further reduced through an iterative procedure solving for the angular and radial velocity of a binary with the same baryon masses and separation but zero eccentricity~\cite{Pfeiffer-Brown-etal:2007,Buonanno:2010yk}. We note that eccentricity is not a uniquely defined quantity for inspiraling binaries. Here, we calculate the eccentricity as in~\cite{Buonanno:2010yk}, i.e. we measure the orbital angular velocity $\Omega(t)$ from the first $2.5$ orbits of the trajectory of the center of mass of the neutron stars, and fit the functional form
\beq
\Omega(t) = A_0 (t_c-t)^{-11/8} + A_1 (t_c-t)^{-13/8}  + B \cos{(\omega t + \phi + C t^2)}
\eeq
with free parameters $A_0, A_1, t_c, \omega,\phi$. By analogy with the eccentricy of Newtonian orbits, we define $e=B/(2\Omega(0)\omega)$. Alternative fits that impose $C=0$ and/or $A_1=0$ can also be used and provide equivalent results for this simple equal mass, non-spinning system.

In this manuscript, all simulations are started $\sim 6$ orbits before merger. We begin by applying the eccentricity reduction procedure to the StellarCollapse equation of state table, with neutron star baryon masses $(1.42M_\odot,1.55M_\odot)$ corresponding to gravitational masses $(1.30M_\odot,1.40M_\odot)$ for neutron stars in isolation. This results in an eccentricity $e\sim 0.003$ in the simulation continued through merger. We then use the same baryon masses, angular velocity, and radial velocity for the Compose equation of state table. We note however that the two tables use different definitions of the baryon mass. Taking the Compose definition $m=N_b m_n$ with $N_b$ the number of baryons and $m_n$ the mass of the neutron, the baryon masses of the StellarCollapse stars are actually $(1.44M_\odot,1.56M_\odot)$. Due to this and other small differences in the equation of state implementations, the Compose simulation has neutron stars with gravitational masses $(1.29M_\odot,1.39M_\odot)$. As gravitational wave emission and orbital evolution are primarily driven by the gravitational masses, this results in slightly different orbits, with residual eccentricity $e\sim 0.01$. While residual eccentricities of that order are commonly used as initial data for binary neutron star mergers in the numerical relativity community, we will see that this resulted in differences in the merger time and dynamical mass ejection beyond the estimated numerical errors. To better differentiate the effect of equation of state implementation and eccentricity, we thus perform an additional simulation with the Compose equation of state table, now keeping the gravitational mass of the neutron stars constant. This brings the eccentricity down to $e\sim 0.004$. We will see that this latter simulation is in better agreement with the simulation performed using the StellarCollapse equation of state table. We note that matching baryon masses for equations of state using different definitions of that quantity was obviously not the right choice to make. The two simulations performed here with stars of the same gravitational mass in fact also have nearly the same baryon mass (see Table~\ref{tab:sim}) when the same reference baryon mass $m_n$ is used consistently for both simulations. This error in our initial choice of matching initial conditions however ends up allowing us to test the outcome of simulations with very similar neutron stars but different eccentricities.

Finally, in order to compare the effect of equation of state implementation with the expected numerical error, we also perform the simulation using the StellarCollapse equation of state table at higher resolution. Specifically, our standard simulation setup is such that our finite difference grid initially has $\sim 91$ grid cells across the diameter of each neutron star, while the high-resolution simulation has $\sim 113$ grid cells across the diameter of each neutron star. These are fairly standard resolutions for pure hydrodynamics or radiation-hydrodynamics simulations, though much coarser than what would be required to perform accurate magnetohydrodynamical simulations of these systems. During the evolution, the grid spacing decreases by as much as $20\%$ as the grid contracts to follow the inspiral of the binary (i.e. the resolution used during the evolution is at most times better than the initial resolution quoted here). The pseudospectral grid used to evolve Einstein's equations uses adaptive mesh refinement, with the number of basis functions in each element chosen to reach a target truncation error in the spectral expansion~\cite{Szilagyi:2009qz,Foucart:2013a}. When going from standard to high resolution, that target error is multiplied by $\left(\Delta x_{\rm HR}/\Delta x_{\rm MR}\right)^3$, with $\Delta x_{\rm MR/HR}$ the grid spacing on the finite difference grid in our standard / high resolution simulation.

We note that we choose these initial conditions because they allow us to study a wide range of physical processes observed in neutron star merger simulations. The mass asymmetry is enough to lead to the production of small amounts of tidal ejecta, $O(10^{-3}M_\odot)$. The system is of low enough mass that the remnant does not immediately collapse to a black hole, leading to the ejection of matter associated with the collision of the two neutron stars; but high enough mass that it is expected to collapse on tens of milliseconds timescales. The same system was for example evolved to black hole collapse in~\cite{Sekiguchi:2016}, with two-moment neutrino transport. In this manuscript, as we lack any microphysics in the simulations, we focus solely on the dynamical ejecta and formation of the early neutron star remnant, leaving the study of black hole formation for simulations including some means of modeling angular momentum transport in the remnant.

\section{Numerical methods}
\label{sec:methods}

We evolve the binary neutron star systems presented here using the Spectral Einstein Code (SpEC)~\cite{SpECwebsite}. SpEC evolves Einstein's equations in the generalized harmonics formalism~\cite{Lindblom2006} on a pseudospectral grid, and the general relativistic equations of hydrodynamics in the Valencia formalism~\cite{Banyuls1997} on a separate finite volume grid. We use a third order Runge-Kutta method for the time discretization, and the WENO5~\cite{Liu1994200,Jiang1996202,Borges} reconstruction and HLL~\cite{HLL} approximate Riemann solver to capture shocks in the fluid evolution. Details of the methods used to evolve binary neutron star systems in SpEC can be found in~\cite{Duez:2008rb,Foucart:2013a}. In these simulations, we ignore magnetic fields and neutrinos.

The initial finite volume grid covers a rectangle $[-45,45]\times[-22.5,22.5]\times[-22.5,22.5]\,{\rm km}$. The initial location of the neutron star centers is at $x_c = \pm 22.5\,{\rm km}$, on the x-axis, and the rectangle rotates with the binary. During tidal disruption, a second level of refinement is created, twice as large and with half the grid resolution. Finally, after merger, we switch to a cubic grid with four levels of refinement. The finest level has sides of length $40\,{\rm km}$, with each coarser level increasing in size by a factor of two. Each refinement level uses $256^3$ cells ($312^3$ at high resolution), for a grid resolution on the finest grid of $156\,{\rm m}$ ($128\,{\rm m}$ at high resolution). We switch from the rectangular to cubical grid as soon as matter outflows at the boundary rises above $0.2M_\odot/s$\footnote{We note that this mass loss occurs over a timescale of a fraction of a millisecond. In all of our simulations, the mass loss at the boundary during that time is less than $10^{-4}M_\odot$}.

\section{Definition of unbound material}
\label{sec:unbound}

We measure the unbound mass following the methods described in~\cite{Foucart:2021ikp}. For ease of comparison with other work, we report results using the geodesic criterion $u_t<-1$, the Bernoulli criterion $hu_t<-h_\infty=-1$ (for our equation of state table), as well as the most advanced model of~\cite{Foucart:2021ikp} that accounts for heating from r-process nucleosynthesis, cooling from neutrino emission during the r-process, and the finite amount of time available for bound material to gain energy from the r-process before it reaches apoastron.

The geodesic criterion typically provides a lower bound on the mass and kinetic energy of the ejecta, while the Bernoulli criterion overestimates the ejecta. Indeed, the geodesic criterion ignores r-process heating and adiabatic expansion of the fluid, while the Bernoulli criterion ignores cooling and the finite time available for r-process heating. The more complex criterion is far from perfect, as it still ignores non-local effects in the evolution of the fluid. In~\cite{Foucart:2021ikp}, we showed that, for the one comparison point available with a full 3D time evolution~\cite{Darbha:2021rqj}, the latter criterion predicted the mass and kinetic energy of the ejecta accurately. It did not however accurately predict its geometry or velocity distribution. We infer from this that our criterion appears to do well at predicting global quantities, but is not an appropriate replacement to long-term hydrodynamics evolution of the ejecta for more detailed properties of the outflows.

We also calculate the mass-weighted root-mean-square average velocity $\langle v_{\rm ej}\rangle$ and mass-weighted average electron fraction $\langle Y_e\rangle$, using the more complex criterion for unbound mass. The average electron fraction is not particularly meaningful for these simulations, however, as we do not include neutrino-matter interactions here and thus ignore what should be the main source of variation of the electron fraction. All simulations find $\langle Y_e\rangle \in [0.035,0.037]$, but the correct electron fraction would certainly be higher when accounting for weak interactions (e.g.~\cite{Sekiguchi:2016} finds $\langle Y_e\rangle\sim 0.27$ for this system).

\section{Results}
\label{sec:results}

\begin{figure}
\includegraphics[width=0.95\columnwidth]{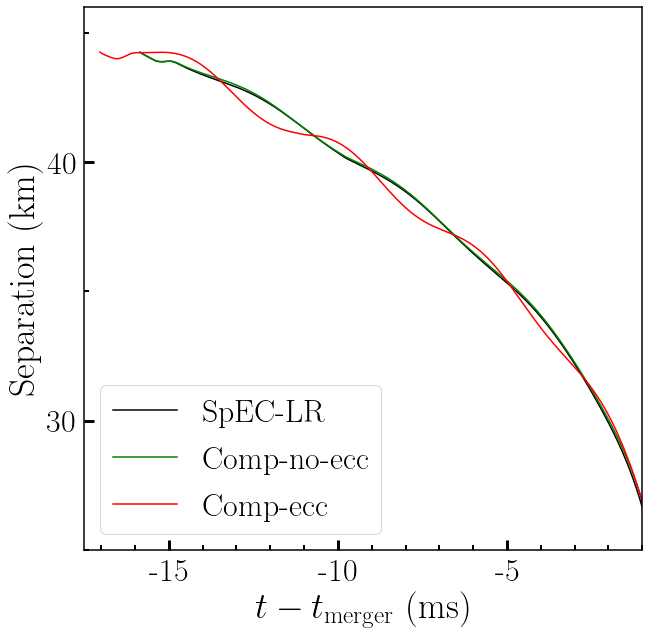}
\caption{Coordinate separation between the center of mass of the neutron star as a function of the time to merger. The high-resolution SpEC simulation would, on this scale, be indinstinguishable from the other two low-eccentricity evolution.}
\label{fig:separation}
\end{figure}

At the beginning of our simulations, the main difference observed between our four cases is due to the residual eccentricity of these systems (or, equivalently, the difference in initial gravitational masses of the neutron stars). The system with $e\sim 0.01$ merges about half an orbit ($1.2\,{\rm ms}$) later than the other simulations. For comparison, the merger times of the other three simulations, defined in SpEC as the time at which the maximum density of the grid reaches $\rho_{\rm max}(t_{\rm merger})=1.03\rho_{\rm max}(t=0)$, are all within $40\,{\rm \mu s}$ of each other. We illustrate these differences on Fig.~\ref{fig:separation}, which shows the coordinate separation between the two neutron stars as a function of the time to merger. 

In all systems, the remnant is a neutron star supported by differential rotation, surrounded by an accretion disk of a few percents of a solar mass. These remants are, in most respects, very similar to each other. We can for example look at the properties of the disk (defined as the mass with $\rho<10^{13}\,{\rm g/cc}$) about $2.5\,{\rm ms}$ post-merger, and find that all simulations show disk masses $M_{\rm disk}\in[4.6-4.8]\times 10^{-2}M_\odot$ with average temperature $T\in [11.5-12.5]\,{\rm MeV}$. The differences in these quantities are largest between the medium and high resolution simulations, indicating that any impact of the eccentricity or equation of state implementation on these quantities is not resolved.

\begin{figure}
\includegraphics[width=0.95\columnwidth]{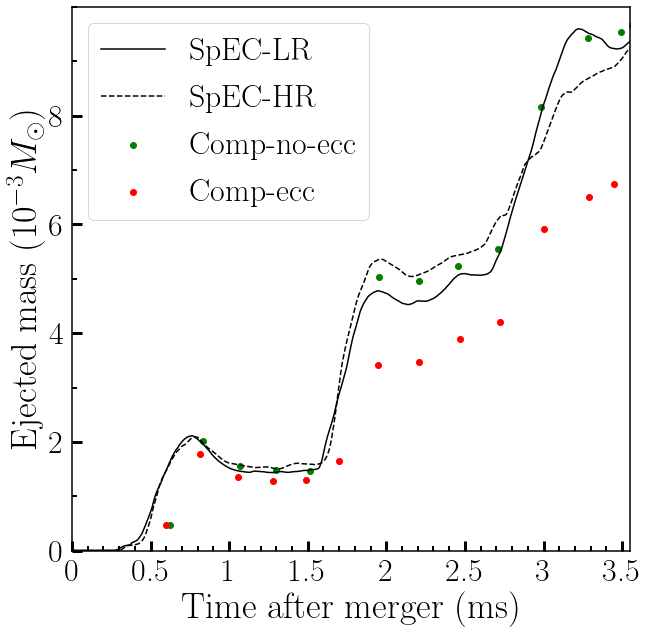}\\
\includegraphics[width=0.95\columnwidth]{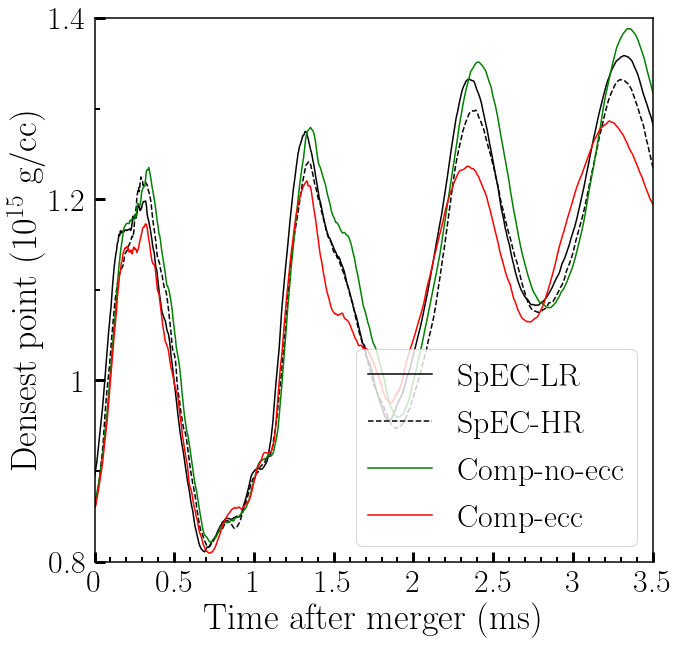}
\caption{{\it Top} Mass marked as unbound on the computational domain for all simulations, using the Bernoulli criteria. The three jumps correspond to the tidal tail and the first two core-bounces. We note that the difference between numerical resolution (solid vs dashed line) is noticeably smaller than the difference between simulations at difference eccentricity (green vs red dots). {\it Bottom}: Densest point on the computational grid as a function of time post-merger, for all simulations.}
\label{fig:unbound}
\end{figure}

\begin{table}
\begin{tabular}{c|c|c|c|c}
{\bf Sim} & $M_{\rm ej}(u_t)$ & $M_{\rm ej}(hu_t)$ & $M_{\rm ej}(rp)$ & $\langle v_{\rm ej}(rp) \rangle$\\
\hline
{\bf Units} & $10^{-3}M_\odot$ & $10^{-3}M_\odot$ & $10^{-3}M_\odot$ & $c$\\\hline
SC-MR &  2.1 & 5.0 & 4.4 & 0.25\\ 
SC-HR &  2.7 & 5.4 & 4.8 & 0.25\\ 
Comp-no-ecc &  2.7 & 5.4 & 4.8 & 0.25\\ 
Comp-ecc & 1.5 & 4.1 & 3.5 & 0.22 \\ 
\end{tabular}
\caption{Unbound mass $M_{\rm ej}$, as measured on the computational grid $2.5\,{\rm ms}$ after merger. The ejected mass is determined using either the geodesic criteria $u_t<-1$, the Bernoulli criteria $hu_t<-1$, or a the more complex method of~\cite{Foucart:2021ikp} approximately accounting for r-process heating and neutrino cooling over the finite time needed for the matter to reach apoastron. We also show the average asymptotic velocity of unbound matter, according to the latter method.}
\label{tab:ej}
\end{table}

\begin{figure}
\includegraphics[width=0.95\columnwidth]{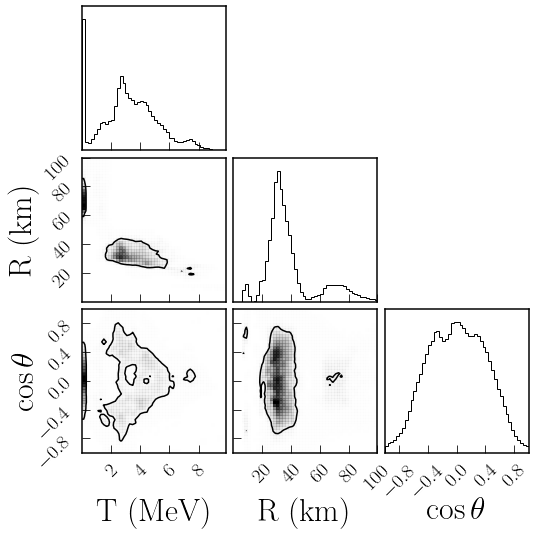}
 \caption{Distribution of the unbound matter shown on Fig.~\ref{fig:unbound}, $2.5\,{\rm ms}$ post-merger and as a function of distance $R$ from the remnant, temperature $T$, and polar angle $\theta$. We can clearly see the two distinct components: the cold, mostly equatorial tidal ejecta about $70\,{\rm km}$ from the remnant, and the hot core-bounce ejecta with broader angular distribution about $30\,{\rm km}$ from the remnant.}
\label{fig:corner}
\end{figure}

The most important output from our simulations that appears impacted by their residual eccentricities is the dynamical ejecta, depicted in Fig.~\ref{fig:unbound} and Fig.~\ref{fig:corner}. In this system, unbound matter is continuously ejected by the remnant, first in the tidal tail, then due to core-bounce and possibly tidal arm instabilities~\cite{Radice:2016gym}. The amount of cold material from the tidal tail marked as unbound stabilizes  about $1\,{\rm ms}$ post-merger at $M_{\rm tail} \in [0.9-1.4]\times 10^{-3}M_\odot$. The effect of eccentricity, of numerical resolution, and of the criterion used to define unbound matter are, at this step, comparable\footnote{The mass quoted here is for the more advanced criterion of~\cite{Foucart:2021ikp}, as will generally be the case here unless otherwise noted. On the other hand, the figure shows results for the Bernoulli criterion, which typically overestimates outflow masses. We use that simple criterion in the figure because the more advanced criterion was only computed as a post-processing step on specific simulation snapshots, and thus we do not have densely sampled data for that criterion. We only show densely sampled results for the StellarCollapse table for the same reason: due to the different definition of the baryon mass in the Compose table, the outflow mass in those simulations was only computed as a post-processing step on a few specific snapshots, shown on the figure.}.  We note that our interpretation of the material as coming from the ``tidal tail'' or ``core-bounce'' is largely based on the very different temperature of the two components, as demonstrated in Fig.~\ref{fig:corner}.

The amount of matter marked as unbound after the first core-bounce is most cleanly measured around $(2-2.5)\,{\rm ms}$ post-merger (see Fig.~\ref{fig:unbound}). As this matter is hotter, the choice of unbound criteria now becomes particularly important -- the geodesic criteria, which ignores the internal energy of the fluid, predicts significantly less ejecta than the other criteria. A summary of the unbound mass for each simulation and each criteria at that time is presented on Table~\ref{tab:ej}. We see that, especially for criteria that do include the internal energy of the fluid, uncertainties in the predicted amount of ejecta are now clearly dominated by residual eccentricity. We see $\sim 25\%-30\%$ differences between the two simulations using the Compose table, with the $e=0.004$ simulation using the Compose table being closer to the $e=0.003$ simulation using the StellarCollapse table than then $e=0.01$ simulation using the Compose table. A clear difference is also observable in the velocity of the ejecta. The average velocity of the ejecta in the three low-eccentricity simulations is $v\in 0.25c$, while it is only $v=0.22c$ for the $e\sim 0.01$ simulation. We will come back to these differences in our discussion of the results below.

At later times, unbound material is more continuously ejected. We can still see the impact of matter ejection from core bounces (at $\sim 3\,{\rm ms}$ post-merger on Fig.~\ref{fig:unbound}), but the exact amount of material flagged as unbound depends more sensitively on the semi-arbitrary definition of merger time used in this work. We also note that we cannot, in these simulations, follow the unbound matter for more than $\sim 3\,{\rm ms}$ after ejection, given the size of our computational grid, so by that time some of the matter counted as unbound on Fig.~\ref{fig:unbound} has left our computational domain (we sample all matter leaving the computational grid, and account for that matter in all masses reported here). The differences in outflow masses between simulations only increase however, with an extra $\sim 0.004\,M_\odot$ of material ejected in low-eccentricity simulations between $2.5-3.5\,{\rm ms}$ post-merger, but only an extra $\sim 0.0025M_\odot$ of material ejected in the $e\sim 0.01$ simulation.

We can get a hint of the physical source of these differences from the bottom panel of Fig.~\ref{fig:unbound}, which shows the maximum density on the computational grid for all four simulations. We see that the $e\sim 0.01$ simulations has weaker core-bounce oscillations, and a lower maximum density at the peak of the oscillations of the post-merger remnant. The $e=0.004$ simulation using the Compose table has slightly larger oscillations of the core than the simulations using the StellarCollapse table. We discuss a potential interpretation of these results in the next section.

\section{Discussion}

In this manuscript, we find that small differences in residual eccentricity ($e\sim 0.01$) $\sim 5$ orbits before merger can have a measurable impact on the observed amount of ejected mass in binary neutron star mergers. The exact process that leads to these differences cannot be rigorously demonstrated from a single set of simulations, yet based on our simulations we can propose a reasonable guess at least. Even a small residual eccentricity can lead the two core to merge in a slightly more ``grazing'' or more ``head-on'' collision, depending on the phase of epicyclic oscillations at the time of merger. In the former case, we would naturally expect the kind of results observed in this manuscript: weaker core-bounce, and as a result less mass ejection. In the latter case, one might however expect the opposite effect. Accordingly, we propose that a small residual eccentricity will have a significant (i.e. numerically resolved) effect on the outflow mass and, likely, the time to black hole formation. On the other hand, the direction in which the residual eccentricity modifies the outflow mass is unknown, as for any given binary the phase of epicyclic oscillations at merger cannot be easily predicted. This interpretation is at least partially supported by the amplitude of the density oscillations observed in the different post-merger remnants simulated in this manuscript, and the fact that the differences in outflow masses are largely due to changes in the amount of hot material ejected during core bounces.

We note that this effect could lead us to predict larger estimated numerical errors in the ejected mass for simulations with $O(1\%)$ residual eccentricity, as simulations performed at different numerical resolutions may merge at different points in their epicyclic oscillations. This effect would of course disappear as resolution is increased and the simulations converge to a well-defined epicyclic phase at merger, but they would converge to a solution for the ejected mass that may have a bias with respect to the $e=0$ answer. This may be an important caveat to existing predictions for mass ejection, as many numerical simulations (including those performed by our own collaboration) have considered residual eccentricities of $O(1\%)$ as acceptable for outflow modeling. We caution however that this interpretation needs to be confirmed with a larger sample of simulations. In particular, we note the possibility that the eccentricity dependence observed here is particularly strong for binaries with total mass close to the prompt collapse threshold (as is the case for the system studied here), for which the core-bounce and early post-merger oscillations are stronger than for lower or higher mass systems.

It is also worth mentioning that the impact of eccentricity in this study is likely comparable to known sources of biases, e.g. the limited microphysics of many simulations~\cite{Nedora:2020qtd}. This should thus be seen as one more source of error to carefully consider when analyzing simulations, though one whose impact may have been underestimated so far.

Finally, we find no clear evidence of disagreement between simulations using the Compose and StellarCollapse versions of the SFHo equation of state -- any actual difference is significantly smaller than the impact of the initial orbital parameters on our results.

\begin{acknowledgments}
F.F. thanks Philipp Moesta and Elias Most for useful discussions during the analysis of these simulations. This research was supported in part by grant NSF PHY-2309135 to the Kavli Institute for Theoretical Physics (KITP).
F.F. gratefully acknowledges support from the Department of Energy, Office of Science, Office of Nuclear Physics, under contract number
DE-AC02-05CH11231 and from the NSF through grant AST-2107932. M.D. gratefully acknowledges support from the NSF through grant PHY-2110287.  M.D. and F.F. gratefully acknowledge support from NASA through grant 80NSSC22K0719. M.S. acknowledges funding from the Sherman Fairchild Foundation
and by NSF Grants No. PHY-1708212, No. PHY-1708213, and No. OAC-1931266
at Caltech.  L.K. acknowledges funding from the Sherman Fairchild Foundation
and by NSF Grants No. PHY-1912081, No. PHY-2207342, and No. OAC-1931280
at Cornell. Computations for this manuscript were performed on the Plasma cluster, a Cray CS500 supercomputer at UNH supported by the NSF MRI program under grant AGS-1919310, and on the Wheeler cluster at Caltech, supported by the Sherman Fairchild Foundation. The authors acknowledge the Texas Advanced Computing Center (TACC) at The University of Texas at Austin and the NSF for providing resources on the Frontera cluster~\cite{10.1145/3311790.3396656} that have contributed to the research results reported within this paper.

\end{acknowledgments}

\bibliography{References/References.bib}

\begin{thebibliography}{35}%
\makeatletter
\providecommand \@ifxundefined [1]{%
 \@ifx{#1\undefined}
}%
\providecommand \@ifnum [1]{%
 \ifnum #1\expandafter \@firstoftwo
 \else \expandafter \@secondoftwo
 \fi
}%
\providecommand \@ifx [1]{%
 \ifx #1\expandafter \@firstoftwo
 \else \expandafter \@secondoftwo
 \fi
}%
\providecommand \natexlab [1]{#1}%
\providecommand \enquote  [1]{``#1''}%
\providecommand \bibnamefont  [1]{#1}%
\providecommand \bibfnamefont [1]{#1}%
\providecommand \citenamefont [1]{#1}%
\providecommand \href@noop [0]{\@secondoftwo}%
\providecommand \href [0]{\begingroup \@sanitize@url \@href}%
\providecommand \@href[1]{\@@startlink{#1}\@@href}%
\providecommand \@@href[1]{\endgroup#1\@@endlink}%
\providecommand \@sanitize@url [0]{\catcode `\\12\catcode `\$12\catcode
  `\&12\catcode `\#12\catcode `\^12\catcode `\_12\catcode `\%12\relax}%
\providecommand \@@startlink[1]{}%
\providecommand \@@endlink[0]{}%
\providecommand \url  [0]{\begingroup\@sanitize@url \@url }%
\providecommand \@url [1]{\endgroup\@href {#1}{\urlprefix }}%
\providecommand \urlprefix  [0]{URL }%
\providecommand \Eprint [0]{\href }%
\providecommand \doibase [0]{https://doi.org/}%
\providecommand \selectlanguage [0]{\@gobble}%
\providecommand \bibinfo  [0]{\@secondoftwo}%
\providecommand \bibfield  [0]{\@secondoftwo}%
\providecommand \translation [1]{[#1]}%
\providecommand \BibitemOpen [0]{}%
\providecommand \bibitemStop [0]{}%
\providecommand \bibitemNoStop [0]{.\EOS\space}%
\providecommand \EOS [0]{\spacefactor3000\relax}%
\providecommand \BibitemShut  [1]{\csname bibitem#1\endcsname}%
\let\auto@bib@innerbib\@empty
\bibitem [{\citenamefont {Abbott}\ \emph {et~al.}(2017)\citenamefont {Abbott}
  \emph {et~al.}}]{TheLIGOScientific:2017qsa}%
  \BibitemOpen
  \bibfield  {author} {\bibinfo {author} {\bibfnamefont {B.~P.}\ \bibnamefont
  {Abbott}} \emph {et~al.} (\bibinfo {collaboration} {Virgo, LIGO
  Scientific}),\ }\bibfield  {title} {\bibinfo {title} {{GW170817: Observation
  of Gravitational Waves from a Binary Neutron Star Inspiral}},\ }\href
  {https://doi.org/10.1103/PhysRevLett.119.161101} {\bibfield  {journal}
  {\bibinfo  {journal} {Phys. Rev. Lett.}\ }\textbf {\bibinfo {volume} {119}},\
  \bibinfo {pages} {161101} (\bibinfo {year} {2017})},\ \Eprint
  {https://arxiv.org/abs/1710.05832} {arXiv:1710.05832 [gr-qc]} \BibitemShut
  {NoStop}%
\bibitem [{\citenamefont {Abbott}\ \emph {et~al.}(2018)\citenamefont {Abbott}
  \emph {et~al.}}]{GW170817-NSRadius}%
  \BibitemOpen
  \bibfield  {author} {\bibinfo {author} {\bibfnamefont {B.}~\bibnamefont
  {Abbott}} \emph {et~al.} (\bibinfo {collaboration} {LIGO Scientific,
  Virgo}),\ }\bibfield  {title} {\bibinfo {title} {{GW170817: Measurements of
  neutron star radii and equation of state}},\ }\href
  {https://doi.org/10.1103/PhysRevLett.121.161101} {\bibfield  {journal}
  {\bibinfo  {journal} {Phys. Rev. Lett.}\ }\textbf {\bibinfo {volume} {121}},\
  \bibinfo {pages} {161101} (\bibinfo {year} {2018})},\ \Eprint
  {https://arxiv.org/abs/1805.11581} {arXiv:1805.11581 [gr-qc]} \BibitemShut
  {NoStop}%
\bibitem [{\citenamefont {Miller}\ \emph {et~al.}(2021)\citenamefont {Miller}
  \emph {et~al.}}]{Miller:2021qha}%
  \BibitemOpen
  \bibfield  {author} {\bibinfo {author} {\bibfnamefont {M.~C.}\ \bibnamefont
  {Miller}} \emph {et~al.},\ }\bibfield  {title} {\bibinfo {title} {{The Radius
  of PSR J0740+6620 from NICER and XMM-Newton Data}},\ }\href
  {https://doi.org/10.3847/2041-8213/ac089b} {\bibfield  {journal} {\bibinfo
  {journal} {Astrophys. J. Lett.}\ }\textbf {\bibinfo {volume} {918}},\
  \bibinfo {pages} {L28} (\bibinfo {year} {2021})},\ \Eprint
  {https://arxiv.org/abs/2105.06979} {arXiv:2105.06979 [astro-ph.HE]}
  \BibitemShut {NoStop}%
\bibitem [{\citenamefont {Raaijmakers}\ \emph {et~al.}(2021)\citenamefont
  {Raaijmakers}, \citenamefont {Greif}, \citenamefont {Hebeler}, \citenamefont
  {Hinderer}, \citenamefont {Nissanke}, \citenamefont {Schwenk}, \citenamefont
  {Riley}, \citenamefont {Watts}, \citenamefont {Lattimer},\ and\ \citenamefont
  {Ho}}]{Raaijmakers:2021uju}%
  \BibitemOpen
  \bibfield  {author} {\bibinfo {author} {\bibfnamefont {G.}~\bibnamefont
  {Raaijmakers}}, \bibinfo {author} {\bibfnamefont {S.~K.}\ \bibnamefont
  {Greif}}, \bibinfo {author} {\bibfnamefont {K.}~\bibnamefont {Hebeler}},
  \bibinfo {author} {\bibfnamefont {T.}~\bibnamefont {Hinderer}}, \bibinfo
  {author} {\bibfnamefont {S.}~\bibnamefont {Nissanke}}, \bibinfo {author}
  {\bibfnamefont {A.}~\bibnamefont {Schwenk}}, \bibinfo {author} {\bibfnamefont
  {T.~E.}\ \bibnamefont {Riley}}, \bibinfo {author} {\bibfnamefont {A.~L.}\
  \bibnamefont {Watts}}, \bibinfo {author} {\bibfnamefont {J.~M.}\ \bibnamefont
  {Lattimer}},\ and\ \bibinfo {author} {\bibfnamefont {W.~C.~G.}\ \bibnamefont
  {Ho}},\ }\bibfield  {title} {\bibinfo {title} {{Constraints on the Dense
  Matter Equation of State and Neutron Star Properties from
  NICER\textquoteright{}s Mass\textendash{}Radius Estimate of PSR J0740+6620
  and Multimessenger Observations}},\ }\href
  {https://doi.org/10.3847/2041-8213/ac089a} {\bibfield  {journal} {\bibinfo
  {journal} {Astrophys. J. Lett.}\ }\textbf {\bibinfo {volume} {918}},\
  \bibinfo {pages} {L29} (\bibinfo {year} {2021})},\ \Eprint
  {https://arxiv.org/abs/2105.06981} {arXiv:2105.06981 [astro-ph.HE]}
  \BibitemShut {NoStop}%
\bibitem [{\citenamefont {Demorest}\ \emph {et~al.}(2010)\citenamefont
  {Demorest}, \citenamefont {Pennucci}, \citenamefont {Ransom}, \citenamefont
  {Roberts},\ and\ \citenamefont {Hessels}}]{Demorest:2010bx}%
  \BibitemOpen
  \bibfield  {author} {\bibinfo {author} {\bibfnamefont {P.}~\bibnamefont
  {Demorest}}, \bibinfo {author} {\bibfnamefont {T.}~\bibnamefont {Pennucci}},
  \bibinfo {author} {\bibfnamefont {S.}~\bibnamefont {Ransom}}, \bibinfo
  {author} {\bibfnamefont {M.}~\bibnamefont {Roberts}},\ and\ \bibinfo {author}
  {\bibfnamefont {J.}~\bibnamefont {Hessels}},\ }\bibfield  {title} {\bibinfo
  {title} {Shapiro delay measurement of a two solar mass neutron star},\ }\href
  {https://doi.org/10.1038/nature09466} {\bibfield  {journal} {\bibinfo
  {journal} {Nature}\ }\textbf {\bibinfo {volume} {467}},\ \bibinfo {pages}
  {1081} (\bibinfo {year} {2010})},\ \Eprint {https://arxiv.org/abs/1010.5788}
  {arXiv:1010.5788 [astro-ph.HE]} \BibitemShut {NoStop}%
\bibitem [{\citenamefont {Horowitz}\ \emph {et~al.}(2014)\citenamefont
  {Horowitz}, \citenamefont {Kumar},\ and\ \citenamefont
  {Michaels}}]{Horowitz:2013wha}%
  \BibitemOpen
  \bibfield  {author} {\bibinfo {author} {\bibfnamefont {C.~J.}\ \bibnamefont
  {Horowitz}}, \bibinfo {author} {\bibfnamefont {K.~S.}\ \bibnamefont
  {Kumar}},\ and\ \bibinfo {author} {\bibfnamefont {R.}~\bibnamefont
  {Michaels}},\ }\bibfield  {title} {\bibinfo {title} {{Electroweak
  Measurements of Neutron Densities in CREX and PREX at JLab, USA}},\ }\href
  {https://doi.org/10.1140/epja/i2014-14048-3} {\bibfield  {journal} {\bibinfo
  {journal} {Eur. Phys. J.}\ }\textbf {\bibinfo {volume} {A50}},\ \bibinfo
  {pages} {48} (\bibinfo {year} {2014})},\ \Eprint
  {https://arxiv.org/abs/1307.3572} {arXiv:1307.3572 [nucl-ex]} \BibitemShut
  {NoStop}%
\bibitem [{\citenamefont {Reed}\ \emph {et~al.}(2021)\citenamefont {Reed},
  \citenamefont {Fattoyev}, \citenamefont {Horowitz},\ and\ \citenamefont
  {Piekarewicz}}]{PhysRevLett.126.172503}%
  \BibitemOpen
  \bibfield  {author} {\bibinfo {author} {\bibfnamefont {B.~T.}\ \bibnamefont
  {Reed}}, \bibinfo {author} {\bibfnamefont {F.~J.}\ \bibnamefont {Fattoyev}},
  \bibinfo {author} {\bibfnamefont {C.~J.}\ \bibnamefont {Horowitz}},\ and\
  \bibinfo {author} {\bibfnamefont {J.}~\bibnamefont {Piekarewicz}},\
  }\bibfield  {title} {\bibinfo {title} {Implications of prex-2 on the equation
  of state of neutron-rich matter},\ }\href
  {https://doi.org/10.1103/PhysRevLett.126.172503} {\bibfield  {journal}
  {\bibinfo  {journal} {Phys. Rev. Lett.}\ }\textbf {\bibinfo {volume} {126}},\
  \bibinfo {pages} {172503} (\bibinfo {year} {2021})}\BibitemShut {NoStop}%
\bibitem [{\citenamefont {Coughlin}\ \emph {et~al.}(2019)\citenamefont
  {Coughlin}, \citenamefont {Dietrich}, \citenamefont {Margalit},\ and\
  \citenamefont {Metzger}}]{Coughlin:2018fis}%
  \BibitemOpen
  \bibfield  {author} {\bibinfo {author} {\bibfnamefont {M.~W.}\ \bibnamefont
  {Coughlin}}, \bibinfo {author} {\bibfnamefont {T.}~\bibnamefont {Dietrich}},
  \bibinfo {author} {\bibfnamefont {B.}~\bibnamefont {Margalit}},\ and\
  \bibinfo {author} {\bibfnamefont {B.~D.}\ \bibnamefont {Metzger}},\
  }\bibfield  {title} {\bibinfo {title} {{Multimessenger Bayesian parameter
  inference of a binary neutron star merger}},\ }\href
  {https://doi.org/10.1093/mnrasl/slz133} {\bibfield  {journal} {\bibinfo
  {journal} {Mon. Not. Roy. Astron. Soc.}\ }\textbf {\bibinfo {volume} {489}},\
  \bibinfo {pages} {L91} (\bibinfo {year} {2019})},\ \Eprint
  {https://arxiv.org/abs/1812.04803} {arXiv:1812.04803 [astro-ph.HE]}
  \BibitemShut {NoStop}%
\bibitem [{\citenamefont {Nedora}\ \emph {et~al.}(2022)\citenamefont {Nedora},
  \citenamefont {Schianchi}, \citenamefont {Bernuzzi}, \citenamefont {Radice},
  \citenamefont {Daszuta}, \citenamefont {Endrizzi}, \citenamefont {Perego},
  \citenamefont {Prakash},\ and\ \citenamefont {Zappa}}]{Nedora:2020qtd}%
  \BibitemOpen
  \bibfield  {author} {\bibinfo {author} {\bibfnamefont {V.}~\bibnamefont
  {Nedora}}, \bibinfo {author} {\bibfnamefont {F.}~\bibnamefont {Schianchi}},
  \bibinfo {author} {\bibfnamefont {S.}~\bibnamefont {Bernuzzi}}, \bibinfo
  {author} {\bibfnamefont {D.}~\bibnamefont {Radice}}, \bibinfo {author}
  {\bibfnamefont {B.}~\bibnamefont {Daszuta}}, \bibinfo {author} {\bibfnamefont
  {A.}~\bibnamefont {Endrizzi}}, \bibinfo {author} {\bibfnamefont
  {A.}~\bibnamefont {Perego}}, \bibinfo {author} {\bibfnamefont
  {A.}~\bibnamefont {Prakash}},\ and\ \bibinfo {author} {\bibfnamefont
  {F.}~\bibnamefont {Zappa}},\ }\bibfield  {title} {\bibinfo {title} {{Mapping
  dynamical ejecta and disk masses from numerical relativity simulations of
  neutron star mergers}},\ }\href {https://doi.org/10.1088/1361-6382/ac35a8}
  {\bibfield  {journal} {\bibinfo  {journal} {Class. Quant. Grav.}\ }\textbf
  {\bibinfo {volume} {39}},\ \bibinfo {pages} {015008} (\bibinfo {year}
  {2022})},\ \Eprint {https://arxiv.org/abs/2011.11110} {arXiv:2011.11110
  [astro-ph.HE]} \BibitemShut {NoStop}%
\bibitem [{\citenamefont {{Barnes}}\ and\ \citenamefont
  {{Kasen}}(2013)}]{2013ApJ...775...18B}%
  \BibitemOpen
  \bibfield  {author} {\bibinfo {author} {\bibfnamefont {J.}~\bibnamefont
  {{Barnes}}}\ and\ \bibinfo {author} {\bibfnamefont {D.}~\bibnamefont
  {{Kasen}}},\ }\bibfield  {title} {\bibinfo {title} {{Effect of a High Opacity
  on the Light Curves of Radioactively Powered Transients from Compact Object
  Mergers}},\ }\href {https://doi.org/10.1088/0004-637X/775/1/18} {\bibfield
  {journal} {\bibinfo  {journal} {Astrophys.\ J.}\ }\textbf {\bibinfo {volume}
  {775}},\ \bibinfo {eid} {18} (\bibinfo {year} {2013})},\ \Eprint
  {https://arxiv.org/abs/1303.5787} {arXiv:1303.5787 [astro-ph.HE]}
  \BibitemShut {NoStop}%
\bibitem [{\citenamefont {East}\ \emph {et~al.}(2016)\citenamefont {East},
  \citenamefont {Paschalidis}, \citenamefont {Pretorius},\ and\ \citenamefont
  {Shapiro}}]{East:2015vix}%
  \BibitemOpen
  \bibfield  {author} {\bibinfo {author} {\bibfnamefont {W.~E.}\ \bibnamefont
  {East}}, \bibinfo {author} {\bibfnamefont {V.}~\bibnamefont {Paschalidis}},
  \bibinfo {author} {\bibfnamefont {F.}~\bibnamefont {Pretorius}},\ and\
  \bibinfo {author} {\bibfnamefont {S.~L.}\ \bibnamefont {Shapiro}},\
  }\bibfield  {title} {\bibinfo {title} {{Relativistic Simulations of Eccentric
  Binary Neutron Star Mergers: One-arm Spiral Instability and Effects of
  Neutron Star Spin}},\ }\href {https://doi.org/10.1103/PhysRevD.93.024011}
  {\bibfield  {journal} {\bibinfo  {journal} {Phys. Rev. D}\ }\textbf {\bibinfo
  {volume} {93}},\ \bibinfo {pages} {024011} (\bibinfo {year} {2016})},\
  \Eprint {https://arxiv.org/abs/1511.01093} {arXiv:1511.01093 [astro-ph.HE]}
  \BibitemShut {NoStop}%
\bibitem [{\citenamefont {{Radice}}\ \emph {et~al.}(2016)\citenamefont
  {{Radice}}, \citenamefont {{Galeazzi}}, \citenamefont {{Lippuner}},
  \citenamefont {{Roberts}}, \citenamefont {{Ott}},\ and\ \citenamefont
  {{Rezzolla}}}]{Radice:2016}%
  \BibitemOpen
  \bibfield  {author} {\bibinfo {author} {\bibfnamefont {D.}~\bibnamefont
  {{Radice}}}, \bibinfo {author} {\bibfnamefont {F.}~\bibnamefont
  {{Galeazzi}}}, \bibinfo {author} {\bibfnamefont {J.}~\bibnamefont
  {{Lippuner}}}, \bibinfo {author} {\bibfnamefont {L.~F.}\ \bibnamefont
  {{Roberts}}}, \bibinfo {author} {\bibfnamefont {C.~D.}\ \bibnamefont
  {{Ott}}},\ and\ \bibinfo {author} {\bibfnamefont {L.}~\bibnamefont
  {{Rezzolla}}},\ }\bibfield  {title} {\bibinfo {title} {{Dynamical Mass
  Ejection from Binary Neutron Star Mergers}},\ }\href
  {https://doi.org/10.1093/mnras/stw1227} {\bibfield  {journal} {\bibinfo
  {journal} {Mon.\ Not.\ Roy.\ Astr.\ Soc.}\ }\textbf {\bibinfo {volume}
  {460}},\ \bibinfo {pages} {3255} (\bibinfo {year} {2016})},\ \Eprint
  {https://arxiv.org/abs/1601.02426} {arXiv:1601.02426 [astro-ph.HE]}
  \BibitemShut {NoStop}%
\bibitem [{\citenamefont {{Steiner}}\ \emph {et~al.}(2013)\citenamefont
  {{Steiner}}, \citenamefont {{Hempel}},\ and\ \citenamefont
  {{Fischer}}}]{Steiner:2012rk}%
  \BibitemOpen
  \bibfield  {author} {\bibinfo {author} {\bibfnamefont {A.~W.}\ \bibnamefont
  {{Steiner}}}, \bibinfo {author} {\bibfnamefont {M.}~\bibnamefont
  {{Hempel}}},\ and\ \bibinfo {author} {\bibfnamefont {T.}~\bibnamefont
  {{Fischer}}},\ }\bibfield  {title} {\bibinfo {title} {{Core-collapse
  Supernova Equations of State Based on Neutron Star Observations}},\ }\href
  {https://doi.org/10.1088/0004-637X/774/1/17} {\bibfield  {journal} {\bibinfo
  {journal} {Astrophys.\ J.}\ }\textbf {\bibinfo {volume} {774}},\ \bibinfo
  {eid} {17} (\bibinfo {year} {2013})},\ \Eprint
  {https://arxiv.org/abs/1207.2184} {arXiv:1207.2184 [astro-ph.SR]}
  \BibitemShut {NoStop}%
\bibitem [{\citenamefont {{O'Connor}}\ and\ \citenamefont
  {{Ott}}(2010)}]{OConnor2010}%
  \BibitemOpen
  \bibfield  {author} {\bibinfo {author} {\bibfnamefont {E.}~\bibnamefont
  {{O'Connor}}}\ and\ \bibinfo {author} {\bibfnamefont {C.~D.}\ \bibnamefont
  {{Ott}}},\ }\bibfield  {title} {\bibinfo {title} {{A New Open-Source Code for
  Spherically-Symmetric Stellar Collapse to Neutron Stars and Black Holes}},\
  }\href {https://doi.org/10.1088/0264-9381/27/11/114103} {\bibfield  {journal}
  {\bibinfo  {journal} {Class.\ Quantum Grav.}\ }\textbf {\bibinfo {volume}
  {27}},\ \bibinfo {eid} {114103} (\bibinfo {year} {2010})},\ \Eprint
  {https://arxiv.org/abs/0912.2393} {arXiv:0912.2393 [astro-ph.HE]}
  \BibitemShut {NoStop}%
\bibitem [{\citenamefont {Typel}\ \emph {et~al.}(2015)\citenamefont {Typel},
  \citenamefont {Oertel},\ and\ \citenamefont {Kl\"ahn}}]{Typel:2013rza}%
  \BibitemOpen
  \bibfield  {author} {\bibinfo {author} {\bibfnamefont {S.}~\bibnamefont
  {Typel}}, \bibinfo {author} {\bibfnamefont {M.}~\bibnamefont {Oertel}},\ and\
  \bibinfo {author} {\bibfnamefont {T.}~\bibnamefont {Kl\"ahn}},\ }\bibfield
  {title} {\bibinfo {title} {{CompOSE CompStar online supernova equations of
  state harmonising the concert of nuclear physics and astrophysics
  compose.obspm.fr}},\ }\href {https://doi.org/10.1134/S1063779615040061}
  {\bibfield  {journal} {\bibinfo  {journal} {Phys. Part. Nucl.}\ }\textbf
  {\bibinfo {volume} {46}},\ \bibinfo {pages} {633} (\bibinfo {year} {2015})},\
  \Eprint {https://arxiv.org/abs/1307.5715} {arXiv:1307.5715 [astro-ph.SR]}
  \BibitemShut {NoStop}%
\bibitem [{\citenamefont {Oertel}\ \emph {et~al.}(2017)\citenamefont {Oertel},
  \citenamefont {Hempel}, \citenamefont {Kl\"ahn},\ and\ \citenamefont
  {Typel}}]{Oertel:2016bki}%
  \BibitemOpen
  \bibfield  {author} {\bibinfo {author} {\bibfnamefont {M.}~\bibnamefont
  {Oertel}}, \bibinfo {author} {\bibfnamefont {M.}~\bibnamefont {Hempel}},
  \bibinfo {author} {\bibfnamefont {T.}~\bibnamefont {Kl\"ahn}},\ and\ \bibinfo
  {author} {\bibfnamefont {S.}~\bibnamefont {Typel}},\ }\bibfield  {title}
  {\bibinfo {title} {{Equations of state for supernovae and compact stars}},\
  }\href {https://doi.org/10.1103/RevModPhys.89.015007} {\bibfield  {journal}
  {\bibinfo  {journal} {Rev. Mod. Phys.}\ }\textbf {\bibinfo {volume} {89}},\
  \bibinfo {pages} {015007} (\bibinfo {year} {2017})},\ \Eprint
  {https://arxiv.org/abs/1610.03361} {arXiv:1610.03361 [astro-ph.HE]}
  \BibitemShut {NoStop}%
\bibitem [{\citenamefont {Pfeiffer}\ \emph {et~al.}(2003)\citenamefont
  {Pfeiffer}, \citenamefont {Kidder}, \citenamefont {Scheel},\ and\
  \citenamefont {Teukolsky}}]{Pfeiffer2003}%
  \BibitemOpen
  \bibfield  {author} {\bibinfo {author} {\bibfnamefont {H.~P.}\ \bibnamefont
  {Pfeiffer}}, \bibinfo {author} {\bibfnamefont {L.~E.}\ \bibnamefont
  {Kidder}}, \bibinfo {author} {\bibfnamefont {M.~A.}\ \bibnamefont {Scheel}},\
  and\ \bibinfo {author} {\bibfnamefont {S.~A.}\ \bibnamefont {Teukolsky}},\
  }\bibfield  {title} {\bibinfo {title} {A multidomain spectral method for
  solving elliptic equations},\ }\href
  {https://doi.org/10.1016/S0010-4655(02)00847-0} {\bibfield  {journal}
  {\bibinfo  {journal} {Comput.\ Phys.\ Commun.}\ }\textbf {\bibinfo {volume}
  {152}},\ \bibinfo {pages} {253} (\bibinfo {year} {2003})},\ \Eprint
  {https://arxiv.org/abs/gr-qc/0202096} {gr-qc/0202096} \BibitemShut {NoStop}%
\bibitem [{\citenamefont {Foucart}\ \emph {et~al.}(2008)\citenamefont
  {Foucart}, \citenamefont {Kidder}, \citenamefont {Pfeiffer},\ and\
  \citenamefont {Teukolsky}}]{FoucartEtAl:2008}%
  \BibitemOpen
  \bibfield  {author} {\bibinfo {author} {\bibfnamefont {F.}~\bibnamefont
  {Foucart}}, \bibinfo {author} {\bibfnamefont {L.~E.}\ \bibnamefont {Kidder}},
  \bibinfo {author} {\bibfnamefont {H.~P.}\ \bibnamefont {Pfeiffer}},\ and\
  \bibinfo {author} {\bibfnamefont {S.~A.}\ \bibnamefont {Teukolsky}},\
  }\bibfield  {title} {\bibinfo {title} {{Initial data for black hole-neutron
  star binaries: A Flexible, high-accuracy spectral method}},\ }\href
  {https://doi.org/10.1103/PhysRevD.77.124051} {\bibfield  {journal} {\bibinfo
  {journal} {Phys. Rev. D}\ }\textbf {\bibinfo {volume} {77}},\ \bibinfo
  {pages} {124051} (\bibinfo {year} {2008})},\ \Eprint
  {https://arxiv.org/abs/0804.3787} {arXiv:0804.3787 [gr-qc]} \BibitemShut
  {NoStop}%
\bibitem [{\citenamefont {Pfeiffer}\ \emph {et~al.}(2007)\citenamefont
  {Pfeiffer}, \citenamefont {Brown}, \citenamefont {Kidder}, \citenamefont
  {Lindblom}, \citenamefont {Lovelace},\ and\ \citenamefont
  {Scheel}}]{Pfeiffer-Brown-etal:2007}%
  \BibitemOpen
  \bibfield  {author} {\bibinfo {author} {\bibfnamefont {H.~P.}\ \bibnamefont
  {Pfeiffer}}, \bibinfo {author} {\bibfnamefont {D.~A.}\ \bibnamefont {Brown}},
  \bibinfo {author} {\bibfnamefont {L.~E.}\ \bibnamefont {Kidder}}, \bibinfo
  {author} {\bibfnamefont {L.}~\bibnamefont {Lindblom}}, \bibinfo {author}
  {\bibfnamefont {G.}~\bibnamefont {Lovelace}},\ and\ \bibinfo {author}
  {\bibfnamefont {M.~A.}\ \bibnamefont {Scheel}},\ }\bibfield  {title}
  {\bibinfo {title} {Reducing orbital eccentricity in binary black hole
  simulations},\ }\href@noop {} {\bibfield  {journal} {\bibinfo  {journal}
  {Class.\ Quantum Grav.}\ }\textbf {\bibinfo {volume} {24}},\ \bibinfo {pages}
  {S59} (\bibinfo {year} {2007})},\ \Eprint
  {https://arxiv.org/abs/gr-qc/0702106} {gr-qc/0702106} \BibitemShut {NoStop}%
\bibitem [{\citenamefont {Buonanno}\ \emph {et~al.}(2011)\citenamefont
  {Buonanno}, \citenamefont {Kidder}, \citenamefont {Mrou\'{e}}, \citenamefont
  {Pfeiffer},\ and\ \citenamefont {Taracchini}}]{Buonanno:2010yk}%
  \BibitemOpen
  \bibfield  {author} {\bibinfo {author} {\bibfnamefont {A.}~\bibnamefont
  {Buonanno}}, \bibinfo {author} {\bibfnamefont {L.~E.}\ \bibnamefont
  {Kidder}}, \bibinfo {author} {\bibfnamefont {A.~H.}\ \bibnamefont
  {Mrou\'{e}}}, \bibinfo {author} {\bibfnamefont {H.~P.}\ \bibnamefont
  {Pfeiffer}},\ and\ \bibinfo {author} {\bibfnamefont {A.}~\bibnamefont
  {Taracchini}},\ }\bibfield  {title} {\bibinfo {title} {{Reducing orbital
  eccentricity of precessing black-hole binaries}},\ }\href
  {https://doi.org/10.1103/PhysRevD.83.104034} {\bibfield  {journal} {\bibinfo
  {journal} {Phys.\ Rev.\ D}\ }\textbf {\bibinfo {volume} {83}},\ \bibinfo
  {pages} {104034} (\bibinfo {year} {2011})},\ \Eprint
  {https://arxiv.org/abs/1012.1549} {arXiv:1012.1549 [gr-qc]} \BibitemShut
  {NoStop}%
\bibitem [{\citenamefont {{Szil{\'a}gyi}}\ \emph {et~al.}(2009)\citenamefont
  {{Szil{\'a}gyi}}, \citenamefont {{Lindblom}},\ and\ \citenamefont
  {{Scheel}}}]{Szilagyi:2009qz}%
  \BibitemOpen
  \bibfield  {author} {\bibinfo {author} {\bibfnamefont {B.}~\bibnamefont
  {{Szil{\'a}gyi}}}, \bibinfo {author} {\bibfnamefont {L.}~\bibnamefont
  {{Lindblom}}},\ and\ \bibinfo {author} {\bibfnamefont {M.~A.}\ \bibnamefont
  {{Scheel}}},\ }\bibfield  {title} {\bibinfo {title} {{Simulations of binary
  black hole mergers using spectral methods}},\ }\href
  {https://doi.org/10.1103/PhysRevD.80.124010} {\bibfield  {journal} {\bibinfo
  {journal} {Phys.\ Rev.\ D}\ }\textbf {\bibinfo {volume} {80}},\ \bibinfo
  {eid} {124010} (\bibinfo {year} {2009})},\ \Eprint
  {https://arxiv.org/abs/0909.3557} {arXiv:0909.3557 [gr-qc]} \BibitemShut
  {NoStop}%
\bibitem [{\citenamefont {{Foucart}}\ \emph {et~al.}(2013)\citenamefont
  {{Foucart}}, \citenamefont {{Deaton}}, \citenamefont {{Duez}}, \citenamefont
  {{Kidder}}, \citenamefont {{MacDonald}}, \citenamefont {{Ott}}, \citenamefont
  {{Pfeiffer}}, \citenamefont {{Scheel}}, \citenamefont {{Szil{\'a}gyi}},\ and\
  \citenamefont {{Teukolsky}}}]{Foucart:2013a}%
  \BibitemOpen
  \bibfield  {author} {\bibinfo {author} {\bibfnamefont {F.}~\bibnamefont
  {{Foucart}}}, \bibinfo {author} {\bibfnamefont {M.~B.}\ \bibnamefont
  {{Deaton}}}, \bibinfo {author} {\bibfnamefont {M.~D.}\ \bibnamefont
  {{Duez}}}, \bibinfo {author} {\bibfnamefont {L.~E.}\ \bibnamefont
  {{Kidder}}}, \bibinfo {author} {\bibfnamefont {I.}~\bibnamefont
  {{MacDonald}}}, \bibinfo {author} {\bibfnamefont {C.~D.}\ \bibnamefont
  {{Ott}}}, \bibinfo {author} {\bibfnamefont {H.~P.}\ \bibnamefont
  {{Pfeiffer}}}, \bibinfo {author} {\bibfnamefont {M.~A.}\ \bibnamefont
  {{Scheel}}}, \bibinfo {author} {\bibfnamefont {B.}~\bibnamefont
  {{Szil{\'a}gyi}}},\ and\ \bibinfo {author} {\bibfnamefont {S.~A.}\
  \bibnamefont {{Teukolsky}}},\ }\bibfield  {title} {\bibinfo {title} {{Black
  hole-neutron star mergers at realistic mass ratios: Equation of state and
  spin orientation effects}},\ }\href@noop {} {\bibfield  {journal} {\bibinfo
  {journal} {Phys.\ Rev.\ D}\ }\textbf {\bibinfo {volume} {87}},\ \bibinfo
  {pages} {084006} (\bibinfo {year} {2013})},\ \Eprint
  {https://arxiv.org/abs/1212.4810} {arXiv:1212.4810 [gr-qc]} \BibitemShut
  {NoStop}%
\bibitem [{\citenamefont {Sekiguchi}\ \emph {et~al.}(2016)\citenamefont
  {Sekiguchi}, \citenamefont {Kiuchi}, \citenamefont {Kyutoku}, \citenamefont
  {Shibata},\ and\ \citenamefont {Taniguchi}}]{Sekiguchi:2016}%
  \BibitemOpen
  \bibfield  {author} {\bibinfo {author} {\bibfnamefont {Y.}~\bibnamefont
  {Sekiguchi}}, \bibinfo {author} {\bibfnamefont {K.}~\bibnamefont {Kiuchi}},
  \bibinfo {author} {\bibfnamefont {K.}~\bibnamefont {Kyutoku}}, \bibinfo
  {author} {\bibfnamefont {M.}~\bibnamefont {Shibata}},\ and\ \bibinfo {author}
  {\bibfnamefont {K.}~\bibnamefont {Taniguchi}},\ }\bibfield  {title} {\bibinfo
  {title} {{Dynamical mass ejection from the merger of asymmetric binary
  neutron stars: Radiation-hydrodynamics study in general relativity}},\ }\href
  {https://doi.org/10.1103/PhysRevD.93.124046} {\bibfield  {journal} {\bibinfo
  {journal} {Phys. Rev.}\ }\textbf {\bibinfo {volume} {D93}},\ \bibinfo {pages}
  {124046} (\bibinfo {year} {2016})},\ \Eprint
  {https://arxiv.org/abs/1603.01918} {arXiv:1603.01918 [astro-ph.HE]}
  \BibitemShut {NoStop}%
\bibitem [{SpE()}]{SpECwebsite}%
  \BibitemOpen
  \href@noop {} {\bibinfo {title} {The {S}pectral {E}instein {C}ode}},\
  \bibinfo {note} {\url{http://www.black-holes.org/SpEC.html}}\BibitemShut
  {NoStop}%
\bibitem [{\citenamefont {Lindblom}\ \emph {et~al.}(2006)\citenamefont
  {Lindblom}, \citenamefont {Scheel}, \citenamefont {Kidder}, \citenamefont
  {Owen},\ and\ \citenamefont {Rinne}}]{Lindblom2006}%
  \BibitemOpen
  \bibfield  {author} {\bibinfo {author} {\bibfnamefont {L.}~\bibnamefont
  {Lindblom}}, \bibinfo {author} {\bibfnamefont {M.~A.}\ \bibnamefont
  {Scheel}}, \bibinfo {author} {\bibfnamefont {L.~E.}\ \bibnamefont {Kidder}},
  \bibinfo {author} {\bibfnamefont {R.}~\bibnamefont {Owen}},\ and\ \bibinfo
  {author} {\bibfnamefont {O.}~\bibnamefont {Rinne}},\ }\bibfield  {title}
  {\bibinfo {title} {A new generalized harmonic evolution system},\ }\href
  {https://doi.org/10.1088/0264-9381/23/16/S09} {\bibfield  {journal} {\bibinfo
   {journal} {Class.\ Quantum Grav.}\ }\textbf {\bibinfo {volume} {23}},\
  \bibinfo {pages} {S447} (\bibinfo {year} {2006})},\ \Eprint
  {https://arxiv.org/abs/gr-qc/0512093} {gr-qc/0512093} \BibitemShut {NoStop}%
\bibitem [{\citenamefont {{Banyuls}}\ \emph {et~al.}(1997)\citenamefont
  {{Banyuls}}, \citenamefont {{Font}}, \citenamefont {{Ib{\'a}{\~n}ez}},
  \citenamefont {{Mart{\'{\i}}}},\ and\ \citenamefont
  {{Miralles}}}]{Banyuls1997}%
  \BibitemOpen
  \bibfield  {author} {\bibinfo {author} {\bibfnamefont {F.}~\bibnamefont
  {{Banyuls}}}, \bibinfo {author} {\bibfnamefont {J.~A.}\ \bibnamefont
  {{Font}}}, \bibinfo {author} {\bibfnamefont {J.~M.}\ \bibnamefont
  {{Ib{\'a}{\~n}ez}}}, \bibinfo {author} {\bibfnamefont {J.~M.}\ \bibnamefont
  {{Mart{\'{\i}}}}},\ and\ \bibinfo {author} {\bibfnamefont {J.~A.}\
  \bibnamefont {{Miralles}}},\ }\bibfield  {title} {\bibinfo {title}
  {{Numerical $\{$3 + 1$\}$ General Relativistic Hydrodynamics: A Local
  Characteristic Approach}},\ }\href {https://doi.org/10.1086/303604}
  {\bibfield  {journal} {\bibinfo  {journal} {Astrophys.\ J.}\ }\textbf
  {\bibinfo {volume} {476}},\ \bibinfo {pages} {221} (\bibinfo {year}
  {1997})}\BibitemShut {NoStop}%
\bibitem [{\citenamefont {Liu}\ \emph {et~al.}(1994)\citenamefont {Liu},
  \citenamefont {Osher},\ and\ \citenamefont {Chan}}]{Liu1994200}%
  \BibitemOpen
  \bibfield  {author} {\bibinfo {author} {\bibfnamefont {X.-D.}\ \bibnamefont
  {Liu}}, \bibinfo {author} {\bibfnamefont {S.}~\bibnamefont {Osher}},\ and\
  \bibinfo {author} {\bibfnamefont {T.}~\bibnamefont {Chan}},\ }\bibfield
  {title} {\bibinfo {title} {Weighted essentially non-oscillatory schemes},\
  }\href {https://doi.org/10.1006/jcph.1994.1187} {\bibfield  {journal}
  {\bibinfo  {journal} {J.\ Comput.\ Phys.}\ }\textbf {\bibinfo {volume}
  {115}},\ \bibinfo {pages} {200 } (\bibinfo {year} {1994})}\BibitemShut
  {NoStop}%
\bibitem [{\citenamefont {Jiang}\ and\ \citenamefont
  {Shu}(1996)}]{Jiang1996202}%
  \BibitemOpen
  \bibfield  {author} {\bibinfo {author} {\bibfnamefont {G.-S.}\ \bibnamefont
  {Jiang}}\ and\ \bibinfo {author} {\bibfnamefont {C.-W.}\ \bibnamefont
  {Shu}},\ }\bibfield  {title} {\bibinfo {title} {Efficient implementation of
  weighted eno schemes},\ }\href {https://doi.org/10.1006/jcph.1996.0130}
  {\bibfield  {journal} {\bibinfo  {journal} {J.\ Comput.\ Phys.}\ }\textbf
  {\bibinfo {volume} {126}},\ \bibinfo {pages} {202 } (\bibinfo {year}
  {1996})}\BibitemShut {NoStop}%
\bibitem [{\citenamefont {Borges}\ \emph {et~al.}(2008)\citenamefont {Borges},
  \citenamefont {Carmona}, \citenamefont {Costa},\ and\ \citenamefont
  {Don}}]{Borges}%
  \BibitemOpen
  \bibfield  {author} {\bibinfo {author} {\bibfnamefont {R.}~\bibnamefont
  {Borges}}, \bibinfo {author} {\bibfnamefont {M.}~\bibnamefont {Carmona}},
  \bibinfo {author} {\bibfnamefont {B.}~\bibnamefont {Costa}},\ and\ \bibinfo
  {author} {\bibfnamefont {W.~S.}\ \bibnamefont {Don}},\ }\bibfield  {title}
  {\bibinfo {title} {An improved weighted essentially non-oscillatory scheme
  for hyperbolic conservation laws},\ }\href
  {https://doi.org/10.1016/j.jcp.2007.11.038} {\bibfield  {journal} {\bibinfo
  {journal} {J.\ Comput.\ Phys.}\ }\textbf {\bibinfo {volume} {227}},\ \bibinfo
  {pages} {3191 } (\bibinfo {year} {2008})}\BibitemShut {NoStop}%
\bibitem [{\citenamefont {Harten}\ \emph {et~al.}(1983)\citenamefont {Harten},
  \citenamefont {Lax},\ and\ \citenamefont {van Leer}}]{HLL}%
  \BibitemOpen
  \bibfield  {author} {\bibinfo {author} {\bibfnamefont {A.}~\bibnamefont
  {Harten}}, \bibinfo {author} {\bibfnamefont {P.~D.}\ \bibnamefont {Lax}},\
  and\ \bibinfo {author} {\bibfnamefont {B.}~\bibnamefont {van Leer}},\
  }\bibfield  {title} {\bibinfo {title} {{On Upstream Differencing and
  Godunov-Type Schemes for Hyperbolic Conservation Laws}},\ }\href
  {https://doi.org/10.1137/1025002} {\bibfield  {journal} {\bibinfo  {journal}
  {SIAM Review}\ }\textbf {\bibinfo {volume} {25}},\ \bibinfo {pages} {35}
  (\bibinfo {year} {1983})}\BibitemShut {NoStop}%
\bibitem [{\citenamefont {Duez}\ \emph {et~al.}(2008)\citenamefont {Duez},
  \citenamefont {Foucart}, \citenamefont {Kidder}, \citenamefont {Pfeiffer},
  \citenamefont {Scheel},\ and\ \citenamefont {Teukolsky}}]{Duez:2008rb}%
  \BibitemOpen
  \bibfield  {author} {\bibinfo {author} {\bibfnamefont {M.~D.}\ \bibnamefont
  {Duez}}, \bibinfo {author} {\bibfnamefont {F.}~\bibnamefont {Foucart}},
  \bibinfo {author} {\bibfnamefont {L.~E.}\ \bibnamefont {Kidder}}, \bibinfo
  {author} {\bibfnamefont {H.~P.}\ \bibnamefont {Pfeiffer}}, \bibinfo {author}
  {\bibfnamefont {M.~A.}\ \bibnamefont {Scheel}},\ and\ \bibinfo {author}
  {\bibfnamefont {S.~A.}\ \bibnamefont {Teukolsky}},\ }\bibfield  {title}
  {\bibinfo {title} {{Evolving black hole-neutron star binaries in general
  relativity using pseudospectral and finite difference methods}},\ }\href
  {https://doi.org/10.1103/PhysRevD.78.104015} {\bibfield  {journal} {\bibinfo
  {journal} {Phys.\ Rev.\ D}\ }\textbf {\bibinfo {volume} {78}},\ \bibinfo
  {eid} {104015} (\bibinfo {year} {2008})},\ \Eprint
  {https://arxiv.org/abs/0809.0002} {arXiv:0809.0002 [gr-qc]} \BibitemShut
  {NoStop}%
\bibitem [{\citenamefont {Foucart}\ \emph {et~al.}(2021)\citenamefont
  {Foucart}, \citenamefont {Moesta}, \citenamefont {Ramirez}, \citenamefont
  {Wright}, \citenamefont {Darbha},\ and\ \citenamefont
  {Kasen}}]{Foucart:2021ikp}%
  \BibitemOpen
  \bibfield  {author} {\bibinfo {author} {\bibfnamefont {F.}~\bibnamefont
  {Foucart}}, \bibinfo {author} {\bibfnamefont {P.}~\bibnamefont {Moesta}},
  \bibinfo {author} {\bibfnamefont {T.}~\bibnamefont {Ramirez}}, \bibinfo
  {author} {\bibfnamefont {A.~J.}\ \bibnamefont {Wright}}, \bibinfo {author}
  {\bibfnamefont {S.}~\bibnamefont {Darbha}},\ and\ \bibinfo {author}
  {\bibfnamefont {D.}~\bibnamefont {Kasen}},\ }\bibfield  {title} {\bibinfo
  {title} {{Estimating outflow masses and velocities in merger simulations:
  Impact of r-process heating and neutrino cooling}},\ }\href
  {https://doi.org/10.1103/PhysRevD.104.123010} {\bibfield  {journal} {\bibinfo
   {journal} {Phys. Rev. D}\ }\textbf {\bibinfo {volume} {104}},\ \bibinfo
  {pages} {123010} (\bibinfo {year} {2021})},\ \Eprint
  {https://arxiv.org/abs/2109.00565} {arXiv:2109.00565 [astro-ph.HE]}
  \BibitemShut {NoStop}%
\bibitem [{\citenamefont {{Darbha}}\ \emph {et~al.}(2021)\citenamefont
  {{Darbha}}, \citenamefont {{Kasen}}, \citenamefont {{Foucart}},\ and\
  \citenamefont {{Price}}}]{Darbha:2021rqj}%
  \BibitemOpen
  \bibfield  {author} {\bibinfo {author} {\bibfnamefont {S.}~\bibnamefont
  {{Darbha}}}, \bibinfo {author} {\bibfnamefont {D.}~\bibnamefont {{Kasen}}},
  \bibinfo {author} {\bibfnamefont {F.}~\bibnamefont {{Foucart}}},\ and\
  \bibinfo {author} {\bibfnamefont {D.~J.}\ \bibnamefont {{Price}}},\
  }\bibfield  {title} {\bibinfo {title} {{Electromagnetic Signatures from the
  Tidal Tail of a Black Hole-Neutron Star Merger}},\ }\href
  {https://doi.org/10.3847/1538-4357/abff5d} {\bibfield  {journal} {\bibinfo
  {journal} {\apj}\ }\textbf {\bibinfo {volume} {915}},\ \bibinfo {eid} {69}
  (\bibinfo {year} {2021})},\ \Eprint {https://arxiv.org/abs/2103.03378}
  {arXiv:2103.03378 [astro-ph.HE]} \BibitemShut {NoStop}%
\bibitem [{\citenamefont {Radice}\ \emph {et~al.}(2016)\citenamefont {Radice},
  \citenamefont {Bernuzzi},\ and\ \citenamefont {Ott}}]{Radice:2016gym}%
  \BibitemOpen
  \bibfield  {author} {\bibinfo {author} {\bibfnamefont {D.}~\bibnamefont
  {Radice}}, \bibinfo {author} {\bibfnamefont {S.}~\bibnamefont {Bernuzzi}},\
  and\ \bibinfo {author} {\bibfnamefont {C.~D.}\ \bibnamefont {Ott}},\
  }\bibfield  {title} {\bibinfo {title} {{One-armed spiral instability in
  neutron star mergers and its detectability in gravitational waves}},\ }\href
  {https://doi.org/10.1103/PhysRevD.94.064011} {\bibfield  {journal} {\bibinfo
  {journal} {Phys. Rev. D}\ }\textbf {\bibinfo {volume} {94}},\ \bibinfo
  {pages} {064011} (\bibinfo {year} {2016})},\ \Eprint
  {https://arxiv.org/abs/1603.05726} {arXiv:1603.05726 [gr-qc]} \BibitemShut
  {NoStop}%
\bibitem [{\citenamefont {Stanzione}\ \emph {et~al.}(2020)\citenamefont
  {Stanzione}, \citenamefont {West}, \citenamefont {Evans}, \citenamefont
  {Minyard}, \citenamefont {Ghattas},\ and\ \citenamefont
  {Panda}}]{10.1145/3311790.3396656}%
  \BibitemOpen
  \bibfield  {author} {\bibinfo {author} {\bibfnamefont {D.}~\bibnamefont
  {Stanzione}}, \bibinfo {author} {\bibfnamefont {J.}~\bibnamefont {West}},
  \bibinfo {author} {\bibfnamefont {R.~T.}\ \bibnamefont {Evans}}, \bibinfo
  {author} {\bibfnamefont {T.}~\bibnamefont {Minyard}}, \bibinfo {author}
  {\bibfnamefont {O.}~\bibnamefont {Ghattas}},\ and\ \bibinfo {author}
  {\bibfnamefont {D.~K.}\ \bibnamefont {Panda}},\ }\bibfield  {title} {\bibinfo
  {title} {Frontera: The evolution of leadership computing at the national
  science foundation},\ }in\ \href {https://doi.org/10.1145/3311790.3396656}
  {\emph {\bibinfo {booktitle} {Practice and Experience in Advanced Research
  Computing}}},\ \bibinfo {series and number} {PEARC '20}\ (\bibinfo
  {publisher} {Association for Computing Machinery},\ \bibinfo {address} {New
  York, NY, USA},\ \bibinfo {year} {2020})\ p.\ \bibinfo {pages}
  {106–111}\BibitemShut {NoStop}%
\end{thebibliography}%

\end{document}